\documentclass[%
 reprint,
 amsmath,amssymb,
 aps,
]{revtex4-1}

\usepackage{graphicx}
\usepackage{dcolumn}
\usepackage{bm}

\usepackage{hyperref}

\usepackage[normalem]{ulem}

\begin{document}

\preprint{APS/123-QED}

\title{Heavy-Fermion Superconductivity in CeAg$_2$Si$_2$\\
--Interplay of Spin and Valence Fluctuations--}

\author{Gernot W. Scheerer}
 \email{gernot.scheerer@unige.ch}
 \affiliation{DQMP - University of Geneva, 1211 Geneva 4, Switzerland.}

\author{Zhi Ren}
 \affiliation{Institute for Natural Sciences, Westlake Institute for Advanced Study, Hangzhou, P. R. China.}

\author{G\'{e}rard Lapertot}
 \affiliation{SPSMS, UMR-E 9001, CEA-INAC/UJF-Grenoble 1, 38054 Grenoble, France.}

\author{Gaston Garbarino}
 \affiliation{European Synchrotron Radiation Facility, 38043 Grenoble Cedex 9, France.}

\author{Didier Jaccard}
 \affiliation{DQMP - University of Geneva, 1211 Geneva 4, Switzerland.}

\date{\today}

\begin{abstract}

We present the pressure-temperature phase diagram of the antiferromagnet CeAg$_2$Si$_2$ established via resistivity and calorimetry measurements under quasi-hydrostatic conditions up to 22.5~GPa.
With increasing pressure, the N\'eel temperature [$T_{\mathrm{N}}(p=0)=8.6$~K] slowly increases up to $T_{\mathrm{N}}=13.4$~K at 9.4~GPa and then vanishes abruptly at the magnetic critical pressure $p_{\mathrm{c}}\sim13$~GPa.
For the first time, heavy fermion superconductivity is observed in CeAg$_2$Si$_2$ .
Superconductivity emerges at $\sim 11$~GPa and persists over roughly 10~GPa.
Partial- and bulk-transition temperatures are highest at $p=16$~GPa, with a maximal $T_{\mathrm{c}^{\rm bulk}}=1.25$~K.
In the pressure region of superconductivity, Kondo and crystal-field splitting energies become comparable and resistivity exhibits clear signatures of a Ce-ion valence crossover.
The crossover line is located at a rapid collapse in resistivity as function of pressure and extrapolates to a valence transition critical endpoint at critical pressure and temperature of $p_{\rm cr}\sim 17$~GPa and $T_{\rm cr}\sim-13~K$, respectively.
Both critical spin and valence fluctuations may build up superconductivity in CeAg$_2$Si$_2$.

\end{abstract}
                       
\maketitle


\section{Introduction}

Superconductivity (SC) in heavy fermion (HF) systems is thought to be mediated by critical fluctuations \cite{miyake86,scalapino86,mathur98,monthoux07,miyake07,stockert11}, while there is no general consensus about the exact nature of the fluctuations.
Especially in the case of SC in the compounds CeCu$_2$Si$_2$ \cite{steglich79,bellarbi84,holmes04}, CeCu$_2$Ge$_2$ \cite{jaccard99}, and CeCu$_2$(Si$_{1-x}$Ge$_{x}$)$_2$ \cite{yuan03}, there are two dominant scenarios.
One considers that the low-pressure SC is mediated by the critical spin fluctuations resulting from the collapse of an antiferromagnetic (AF) phase at $p_{\mathrm{c}}$ \cite{stockert11,arndt11}.
The other, mainly defended by Prof. K. Miyake and coworkers, is that critical valence fluctuations (CVF) of a valence crossover (VCO) at $p_{\mathrm{v}}$ play the dominant role for the high-pressure SC, when $T_{\mathrm{c}}$ is optimal close to $p_{\mathrm{v}}$ \cite{holmes07,watanabe11,seyfarth12}.
Another noteworthy case is the most recently discovered HF superconductor CeAu$_2$Si$_2$\cite{ren14}. Its phase diagram is markedly different from all previous cases and --showing also clear features of the CVF picture \cite{ren14,ren15,ren16,scheerer17} -- revives the debate about the current understanding of superconductivity and magnetism in heavy fermion systems.
Naturally, the next step is to the search for SC in the isoelectronic and isostructural compound CeAg$_2$Si$_2$.

CeAg$_2$Si$_2$ has a tetragonal ThCr$_2$Si$_2$ structure with the space group $I4/mmm$ ($D^{17}_{4\mathrm{h}}$) \cite{grier84}.
The Ce ion is nearly trivalent and its 4f $J=\frac{5}{2}$ multiplet is split into 3 doublets by the crystal electric field (CEF) with the first and second exited doublets at 8.8 and 18.0~meV, respectively \cite{severig89b}.
The system exhibits weak Kondo-lattice resistivity and thermopower behavior \cite{garde94} and the estimated Kondo temperature is $T_{\rm K}=1.7$~K \cite{severig89b}.
CeAg$_2$Si$_2$ undergoes an AF transition at the N\'eel temperature $T_{\mathrm{N}}=8.6$~K \cite{book82,thompson86,palstra86}.
The magnetic order is either a spin density wave or a square-wave structure both with spins aligned along the \textit{a}-axis \cite{grier84}.
A previous pressure study (limited to 1.5~GPa) indicates that $T_{\mathrm{N}}$ increases linearly with increasing pressure \cite{thompson86}.

We scanned the pressure-temperature ($p$-$T$) phase diagram of high quality CeAg$_2$Si$_2$ crystals via resistivity and ac heat capacity measurements, and discovered a large dome of SC centered around $p=16$~GPa with a maximum critical temperature of $T_{\mathrm{c}}=1.25$~K.
With increasing pressure, $T_{\rm{N}}$ increases up to a maximum $T_{\rm{N}}=13.4$~K at $p\sim9.4$~GPa, then decreases and vanishes rapidly at the magnetic critical pressure $p_{\rm{c}}\sim13$~GPa.
SC emerges close to $p_{\rm{c}}$ and persists over roughly 10~GPa.
An effective mass of $\sim190\times$ the free electron mass is deduced from the large initial slope of the upper critical field, indicating that SC is build up by heavy quasiparticles.
Properties of the normal-state resistivity (residual resistivity $\rho_0$, power-law coefficient $A$ and exponent $n$) are extracted under an applied magnetic field.
Strongly enhanced scattering rates occur around $p_{\rm{c}}$, indicating enhanced spin fluctuations. A drastic collapse of the coefficient $A$ in the pressure region of SC is ascribed to the rapid delocalization of the Ce 4f electrons.
Resistivity exhibits a scaling behavior suggesting the presence of a VCO, similar to that in CeCu$_2$Si$_2$ \cite{seyfarth12} and CeAu$_2$Si$_2$ \cite{ren14}.
The VCO arises from the critical end point (CEP) of an underlying, putative first-order-valence transition at negative temperature.
The resulting critical pressure and temperature of the CEP are $p_{\rm{cr}}\sim17$~GPa and $T_{\rm{cr}}\sim-13$~K, respectively.
We discuss the possibility that SC in CeAg$_2$Si$_2$ is driven by critical valence fluctuations.

\section{Experiment}

The experiment was done with CeAg$_2$Si$_2$ single crystals (see Ref. \cite{ren14} for details on crystal-growth method).
The data were obtained in two different Bridgman-type pressure cells, one with sintered-diamond anvils ($p^{\rm max}=22.5$~GPa) and the other with tungsten-carbide anvils ($p^{\rm max}=10$~GPa).
Both cells were closed with a pyrophyllite gasket and filled with steatite as soft-solid pressure medium.
Temperature and magnetic field were controlled by a standard dilution fridge equipped with a superconducting magnet coil ($\mu H^{\rm max}=8.5$~T).
Rod shaped samples were cut from the same batch with a slow-cut diamond saw ($\approx20\times50\times600$~$\mu$m$^3$, sample length along the basal plane).
The samples residual resistivity and RRR at ambient pressure are 5.5~$\mu\Omega\cdot$cm and 4.55 respectively.
The magnetic field was applied along the \textit{c}-axis.
The dc resistivity was measured with linear four-point contacts and the ac heat capacity via a system of local heater and thermocouple (see Refs. \cite{link96,holmes04,ren14} for technical details). Heat capacity was measured in the tungsten-carbide anvil cell for pressures $2\leq p\leq 10$~GPa.
Pressure was determined from the resistive superconducting transition temperature of a lead strip. The pressure gradient along the sample estimated from the transition width was $\approx0.5$~GPa at low pressure and increased progressively up to $\approx1.4$~GPa at maximum pressure.

Additionally, we studied the compressibility of CeAg$_2$Si$_2$ at ambient temperature via X-ray scattering (GSAS and Crysalis at ESRF, data not shown here).
An anomaly is found in the compressibility at $\sim20$~GPa, which may be related to a valence instability.


\section{Results}

\begin{figure}[htbp]
\centering
\includegraphics[width=0.9\linewidth]{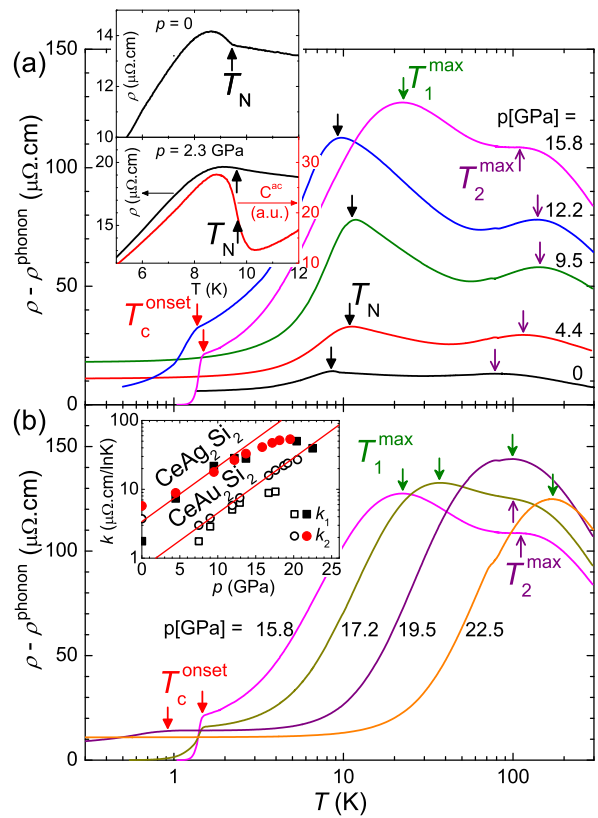}
\caption{
(a,b) Resistivity $\rho-\rho_{\rm{ph}}$ versus temperature $T$ (log-scale) of CeAg$_2$Si$_2$ at selected pressures, where $\rho_{\rm{ph}}$ is the phonon-term derived from LaPd$_2$Si$_2$.
Arrows indicate the N\'eel temperature $T_{\mathrm{N}}$, onset temperature $T_{\rm{c}}^{\rm{onset}}$ of SC, and the characteristic temperatures $T_{1}^{\mathrm{max}}$ and $T_{2}^{\mathrm{max}}$ (see text). Bulk SC occurs at the $\rho=0$ criteria.
Inset of (a): $\rho(T)$ at zero pressure and $\rho(T)$ and $C^{\rm{ac}}(T)$ at 2.3~GPa around $T_{\mathrm{N}}$.
Inset of (b): -ln$T$ slopes $k_1$ and $k_2$ extracted via linear fits on $\rho($ln$T)$ above $T_{1}^{\mathrm{max}}$ and $T_{2}^{\mathrm{max}}$, respectively. $k_1$ and $k_2$ of CeAu$_2$Si$_2$ from Ref. \cite{ren16} is added for comparison. The solid lines are a guide to the eyes.
}
\label{intro}
\end{figure}

Figure 1 presents the electronic resistivity $\rho-\rho_{\rm{ph}}$ versus temperature $T$ of CeAg$_2$Si$_2$ measured at pressures $p$ up to 22.5~GPa, where $\rho_{\rm{ph}}$ is the pressure-independent phonon term derived from LaPd$_2$Si$_2$ \cite{mo89}.
Overall temperature and pressure dependences of resistivity are typical for a Ce-based Kondo lattice \cite{jaccard99}. Characteristic are the anomaly at the N\'eel temperature $T_{\rm{N}}$, the two broad maxima at $T_1^{\rm{max}}$ and $T_2^{\rm{max}}$, and the resistivity drop at the superconducting transition temperature $T_{\rm{c}}$.
The amplitude of $\rho$ at $T_1^{\rm{max}}$ increases with increasing pressure due to the increasing c-f coupling $J_{\rm K}$.

The signature of the AF transition in $\rho$ at $T_{\rm{N}}$ is sharp at zero pressure but gets more and more blurred with increasing pressure [see inset of Fig. 1(a)].
However, the pressure induced increase of $T_{\rm{N}}$ is clearly established by the anomaly in the ac heat capacity $C^{\rm ac}$ [see inset of Fig. 1(a)].
Missing $C^{\rm ac}$ data above 10~GPa, the rapid vanishing of $T_{\rm{N}}$ between 12.2 and 13.7~GPa is established by resistivity only.
Below $T_N$, $\rho$ drops rapidly with decreasing temperature and is governed by electron-magnon scattering.
The maxima at $T_1^{\rm{max}}$ and $T_2^{\rm{max}}$ are due to the crossover from coherent to incoherent Kondo scattering on the Ce 4f-ground state and excited CEF levels, respectively \cite{cornut72}.
At low pressure, $T_1^{\rm{max}}<T_{\rm{N}}$ and the resistivity maximum at $T_1^{\rm{max}}$ is cut by the onset of coherent electron-magnon scattering.
$T_1^{\rm{max}}$ increases rapidly with increasing pressure, while $T_2^{\rm{max}}$ is rather $p$-independent.
The resistivity maxima at $T_1^{\rm{max}}$ and $T_2^{\rm{max}}$ merge at $\sim 19$~GPa, i.e. near the pressure of optimal SC [see also Fig. 2(a)], which is actually a hallmark of Ce-based HF superconductors \cite{jaccard85,jaccard99,wilhelm00,wilhelm02,demuer02,ren14,noteCeRhIn5} and indicates that Kondo and CEF splitting energies become comparable in this pressure region \cite{nishida06}.
Above $\sim 19$~GPa, a single maximum remains due to the dominating contribution of $T_1^{\rm{max}}$.
In the incoherent scattering regimes above $T_1^{\rm{max}}$ and $T_2^{\rm{max}}$, the resistivity decreases as $-k_{\rm i}\mathrm{ln} T$ with the logarithmic slopes $k_1$ and $k_2$, respectively.
The pressure dependence of $k_1$ and $k_2$ [inset of Fig. 1(b)] is similar to that found in CeAu$_2$Si$_2$ \cite{ren16} and points to a constant ratio $k_2/k_1$ as expected from theory \cite{cornut72}. However, one expects $k_1<k_2$, which hints to an underestimation of the phonon contribution $\rho_{\rm{ph}}$.
The saturation of $k_1$ and $k_2$ at high pressures is an artifact due to the convergence of $T_1^{\rm{max}}$ and $T_2^{\rm{max}}$ and the limited temperature scale, respectively.
The significant increase of the $-\mathrm{ln} T$ slopes indicates the pressure induced enhancement of the Kondo temperature $T_{\rm{K}}$ by roughly one order of magnitude over the investigated pressure range \cite{cornut72}.

Partial SC sets in at $T_{\rm{c}}^{\rm{onset}}$ and the transition completes with $\rho=0$, where bulk SC is presumably established \cite{note1}.
Optimal SC with  with $T_{\rm{c}}^{\rm{bulk}}=1.25$~K and $T_{\rm{c}}^{\rm{onset}}=1.57$~K occurs at $p=15.8$~GPa.
At the same pressure, scans of the resistive transition under magnetic field yield the SC upper critical field $H_{\mathrm{c2}}=5.75$~T and the initial slope
$H_{\mathrm{c2}}'=d H_{\mathrm{c2}}/d T|_{T\rightarrow T_{\mathrm{c}}}\approx -11$~T/K for the $\rho=0$ criteria.
The onset criteria yields $H_{\mathrm{c2}}^{\mathrm{onset}}\approx7.5$~T and $H_{\mathrm{c2}}'^{\mathrm{onset}}\approx25$~T/K.
The $H_{\mathrm{c2}}'$ value gives a clean-limit effective mass of $m^*\sim190$~$m_0$ \cite{orlando79}, where $m_0$ is the free-electron mass. Thus, SC in CeAg$_2$Si$_2$ is build up by HF quasiparticles.

Figure 2(a) presents the $p$-$T$ phase diagram of CeAg$_2$Si$_2$ resulting from our resistivity and heat capacity measurements.
Again, the overall appearance is the one of a Ce-based HF system. A low-pressure AF phase vanishes at the critical pressure $p_{\rm{c}}$ resulting in a paramagnetic ground state above $p_{\rm{c}}$ and an intermediate valence regime is reached at still higher pressure.
All is controlled by the increasing Kondo scale $T_{\rm K}$, which is initially of the order of 2~K and increases rapidly by more than one order of magnitude, as indicated by the characteristic temperature $T_1^{\rm{max}}$.

Peculiarity in CeAg$_2$Si$_2$ are the details of the magnetic and superconducting phases.
The N\'eel temperature $T_{\rm{N}}$ ($=8.6$~K at $p=0$) increases with increasing pressure up to a maximum $T_{\rm{N}}=13.4$~K at $p\sim9.4$~GPa and suddenly vanishes at $p_{\rm{c}}$ between 12.2 and 13.7~GPa. Note that $T_{\rm{N}}$ drops from about 10~K to zero over a narrow pressure interval of less than 1.5~GPa. Considering a $p$ gradient along the sample of $\sim1$~GPa, the collapse is much more rapid than usually observed in Ce-based HF compounds.

Partial SC emerges at around 11~GP just before the collapse of the AF order at $p_{\rm{c}}$ and spans over a remarkably wide pressure range of roughly 10~GPa.
Bulk SC ($\rho=0$) emerges around 13.5~GPa and vanishes around 17.5~GPa, which still is a broad pressure range of $\sim4$~GPa.
The dome of SC culminates at $p=15.8$~GPa with $T_{\rm{c}}^{\rm{bulk}}=1.25$~K and $T_{\rm{c}}^{\rm{onset}}=1.57$~K.
One can expect to find a higher $T_{\rm{c}}$ and a much larger area of bulk SC in samples of better quality \cite{ren15}.
The overlap of partial SC and AF phase of roughly 2~GPa is small compared to the total area of SC and optimal bulk SC occurs roughly 3~GPa above the AF collapse at $p_{\rm{c}}$.
For comparison, there is a similar small overlap in CeCu$_2$(Si/Ge)$_2$ \cite{jaccard99}, where the maximum $T_{\rm{c}}$ occurs at $p\sim p_{\rm{c}} + 4.5$~GPa, while CeAu$_2$Si$_2$ \cite{ren14} exhibits is a huge overlap ($\sim12$~GPa) of SC with the magnetic phase, where latter vanishes just at the maximum $T_{\rm{c}}$.
Common features are the relatively high maximal $T_{\rm{c}}$ ($\geq1.25$~K) and the very broad domes of SC ($\Delta p>7$~GPa), which distinguish the CeCu$_2$Si$_2$-HF family.

\begin{figure}[htbp]
\centering
\includegraphics[width=1\linewidth]{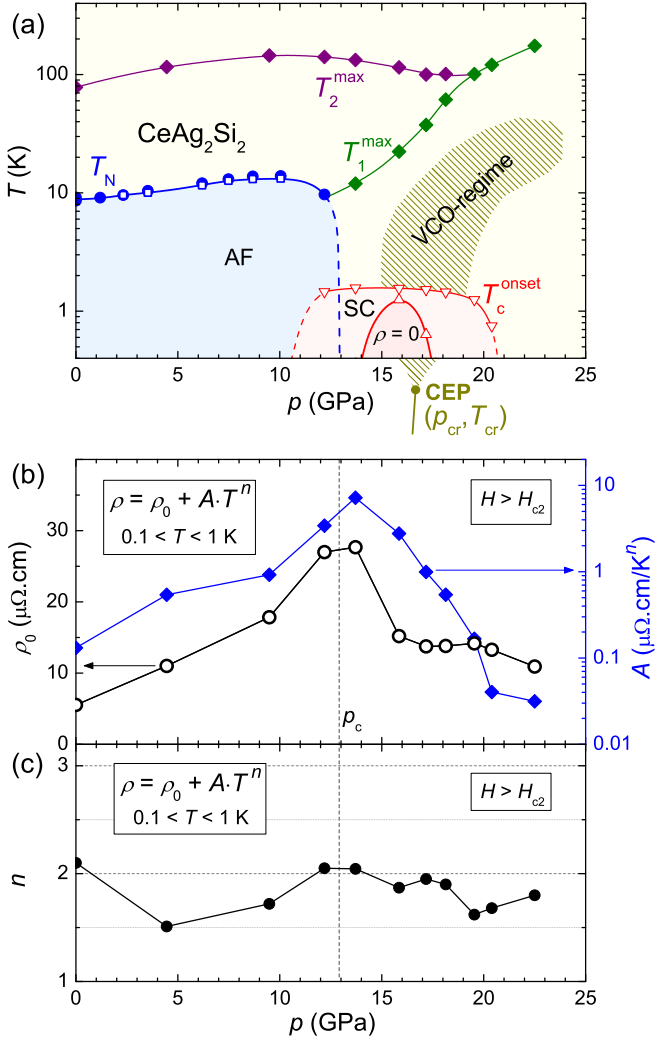}
\caption{
(a) $p$-$T$ phase diagram of CeAg$_2$Si$_2$ resulting from resistivity and ac heat capacity measurements in quasi-hydrostatic conditions. 
$T_1^{\rm{max}}$, $T_2^{\rm{max}}$, and $T_{\rm{c}}$ are based on resistivity. $T_{\rm{N}}$ is based on resistivity (full circles) and ac heat capacity (empty squares).
(b) and (c) Pressure dependences of the fitting parameters of the power law $\rho(T)=\rho_0+A\cdot T^n$ extracted from the the normal-state resistivity ($H=8$~T $>H_{\mathrm{c2}}^{\rm{onset}}$) for $0.1<T<1$~K.
}
\label{intro}
\end{figure}

Let us now discuss the properties of the normal-state resistivity measured under a magnetic field $H=8$~T $>H_{\mathrm{c2}}^{\rm{onset}}$.
At lowest temperatures ($T<\frac{1}{10}T_{\rm{N}}$, $T<\frac{1}{10}T_1^{\rm{max}}$), the resistivity $\rho(T)$ is well described by a simple power law $\rho=\rho_0+A\cdot T^n$.
Figure 2 presents the pressure dependences of the residual resistivity $\rho_0$, the coefficient $A$, and the exponent $n$.
Both $\rho_0$ and $A$ increase with increasing pressure by a factor of $\approx5$ and $\approx55$, respectively, up to a maximum at $p=13.7$~GPa close to the magnetic $p_{\rm{c}}$.
Above 13.7~GPa, $\rho_0$ decreases by a factor of 2 and stabilizes on a shoulder at $p\sim19$~GPa.
The coefficient $A$ decreases by more than 2 orders of magnitude from 13.7~GPa up to $p^{\rm max}=22.5$~GPa.
The peak in $\rho_0$ around $p_{\rm{c}}$ is ascribed to enhanced impurity scattering presumably due to enhanced spin \cite{miyake02} or charge \cite{hattori10} fluctuations, and the shoulder around 19~GPa may be due to a VCO \cite{miyake02a}.
The 55 fold increase in $A$ may be attributed to two different phenomena, which are the recovering of paramagnetic c-f Kondo scattering through the collapse of magnetic order \cite{scheerer17} and the enhancement of the quasiparticles effective mass $m^*\propto\sqrt{A}$ due to critical fluctuations \cite{flouquet04}.
The drastic collapse in $A$ above 13.5~GPa results from the rapid delocalization of the Ce 4f electrons, when pressure drives the system from trivalent to intermediate valence regime \cite{holmes04,seyfarth12}.
Identical observations in CeCu2Si2 \cite{vargoz98,holmes04}, CeCu2Ge2 \cite{jaccard99}, and CeAu2Si2 \cite{ren14,scheerer17} are attributed to a pressure induced VCO at $p_{\rm{v}}$ close to the pressure of optimal SC.

From residual electron-magnon scattering in the AF state ($T<<T_{\rm{N}}$), one expects a Fermi-liquid like $n\approx 2$, which is the case at zero pressure and close to $p_{\rm{c}}$.
At intermediate pressures $\sim5$~GPa, values of $n$ smaller than 2 may be due to a magnetic structure effect, as observed in the parent compounds \cite{knebel96,vargozThesis,scheerer17}.
In the paramagnetic state above $p_{\rm{c}}$, $n$ is scattered between $\sim1.6$ -- 2 with a minimum at 19.5~GPa.

\begin{figure}[htbp]
\centering
\includegraphics[width=0.925\linewidth]{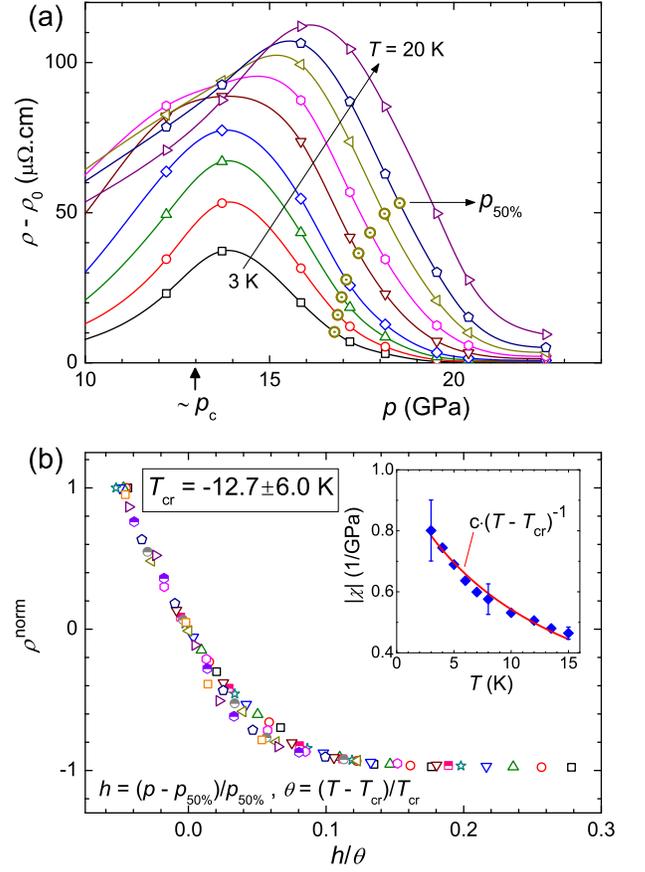}
\caption{
(a) Isotherms of $\rho-\rho_0$ at temperatures between 3 and 20~K.
The pressure $p_{50\%}$ is defined at the 50\% drop of $\rho-\rho_0$ compared to an initial value at $p_{\rm{in}}=15.85$~GPa. The spline lines are guides to the eyes.
(b) Normalized resistivity isotherms $\rho^{\rm{norm}}$ versus the generalized distance from the CEP $h/\theta$ (see text for details) for temperatures from 3 to 22.5~K.
Inset: Temperature dependence of the slope $\chi$ (see text for details). The value $T_{\rm{cr}}=-12.7$~K is extracted via a fit with $\chi=c\cdot(T-T_{\rm{cr}})^{-1}$ (red line).
The estimated error on $T_{\rm{cr}}$ is $\pm 6.0$~K.
}
\label{intro}
\end{figure}

The existence of the putative VCO can be corroborated by a resistivity scaling analysis, which was first elaborated by Seyfarth \textit{et al}. \cite{seyfarth12} on data from CeCu$_2$Si$_2$.
Before applying the same method on the present $\rho$ data, we recall that theory based on the extension of the periodic Andersen model by the supplementary term $U_{\rm fc} \sum^N_{i=1} n_i^f n_i^c$, where $U_{\rm fc}$ is the Coulomb repulsion between f and conduction electrons, predicts a pressure induced VCO for small values of $U_{\rm fc}$ and predicts CVF mediated Cooper pairing in proximity of the VCO \cite{onishi00,watanabe06}.
In CeCu$_2$Si$_2$, such a VCO with a critical end point (CEP) at slightly negative critical temperature $T_{\rm{cr}}\sim -8$~K was found at $p_{\rm v}\sim 4$~GPa, i.e., just at the pressure of optimal SC \cite{seyfarth12}.

The scaling analysis \cite{seyfarth12} starts with plotting the resistivity isotherms $\rho^*=\rho-\rho_0$ at different temperatures, as shown in Fig. 3(a).
After a maximum at $p\sim15$~GPa, $\rho^*$ decreases by one to two orders of magnitude with increasing pressure.
As the collapse of the $A$ coefficient in the same $p$ region, the decrease in $\rho^*$ is a direct consequence of the rapid delocalization of the f electrons through the VCO.
The absence of any discontinuity in $\rho^*(p)$ down to lowest temperatures confirms the crossover nature of the valence instability.
At lowest temperature the drop of $\rho^*(p)$ is steepest at $\approx16$~GPa, indicating that the center of the VCO occurs close to optimal SC [see Fig. 2(a)].

Following \cite{seyfarth12}, we define a normalized resistivity $\rho^{\rm norm}(p)=\frac{\rho^*(p)-\rho^*(p_{50\%})}{\rho^*(p_{50\%})}$, where $p_{50\%}$ is defined by the 50\% drop of $\rho^*(p)$ compared to an initial value at $p_{\rm in}=15.8$~GPa (fixed for all temperatures). $p_{\rm in}$ marks the onset of the decrease of $\rho^*(p)$ at intermediate temperatures and $p_{\rm in}>p_{\rm c}$.
The normalization is necessary to separate the effect of the f electron delocalization from the temperature dependence.
The steepness of the resistivity collapse defined as $\chi=|\mathrm{d} \rho^{\rm norm} /\mathrm{d} p|_{p_{50\%}}$ diverges when $T\rightarrow T_{\rm{cr}}$ [inset of Fig. 3(b)].
A fit with the well established temperature dependence $\chi\propto(T-T_{\rm{cr}})^{-1}$ \cite{seyfarth12,ren14,ren15} yields $T_{\rm{cr}}=-12.7\pm 6.0$~K.
Note that the analysis is limited to $T\leq15$~K, which corresponds to a small fraction to of the first CEF-splitting energy.
Now, one can calculate the general distance $h/\theta$ from the CEP, where $h=(p-p_{50\%})/p_{50\%}$ and $\theta=(T-T_{\rm{cr}})/|T_{\rm{cr}}|$.
Figure 3(b) shows that all curves $\rho^{\rm norm}$ versus $h/\theta$ collapse on a single scaling function.
The resistivity behavior of CeAg$_2$Si$_2$ clearly indicates a pressure-induced VCO, which arises from the CEP of an underlying first-order-valence transition at critical negative temperature $T_{\rm{cr}}\sim-13$~K and at critical pressure $p_{\rm{cr}}$ of roughly 17~GPa.

\section{Discussion}

In CeAg$_2$Si$_2$ and in its isovalent and isostructural parent compounds CeCu$_2$Si$_2$, CeCu$_2$Ge$_2$ and CeAu$_2$Si$_2$ resistivity behaves very similar, and especially the merge of $T_1^{\rm max}$ and $T_2^{\rm max}$, the rapid collapse of $A$ and $\rho^*$ as function of pressure, and the resistivity-scaling behavior are recurring signatures ascribed to a valence instability \cite{jaccard99,holmes04,miyake07,seyfarth12,ren14,ren16}.
Following the scaling analysis of Ref. \cite{seyfarth12}, systematically, a VCO is found at pressures close to the pressure of optimal SC \cite{seyfarth12,ren14}.
Therefore, one has to consider valence fluctuations as plausible superconducting mechanism \cite{onishi00,watanabe06}.
In such a scenario, a less higher $T_{\rm{c}}=1.25$~K in CeAg$_2$Si$_2$ compared to $T_{\rm{c}}\approx2.5$~K in CeCu$_2$Si$_2$ and CeAu$_2$Si$_2$ may be due to a moderately high residual scattering \cite{okada11} and weaker valence fluctuations at finite temperatures from a more negative CEP \cite{onishi00}.
Later may also explain that CeAg$_2$Si$_2$ exhibits only a minumum $n=1.6$ and a weak shoulder in $\rho_0$ in the VCO regime, while CVF theory predicts non-fermi liquid behavior with $n=1$ \cite{holmes04} and a peak in $\rho_0$ around $p_{\rm{v}}$ \cite{miyake02a} as observed in CeCu$_2$Si$_2$ \cite{holmes04}.

Spin fluctuations are believed to be the canonical superconducting mechanism near a magnetic quantum critical point (QCP) in Ce-based HF compounds \cite{monthoux07}.
Indeed, SC emerges close to the collapse of AF order in CeAg$_2$Si$_2$.
However let us mention some points, which are not in favor of the spin fluctuation scenario.

(i) The rapid collapse of magnetism and a power law exponent $n\approx 2$ around $p_{\rm{c}}$ may be interpret as the absence of a magnetic QCP.
It has been shown theoretically \cite{watanabeJPSJ10a,watanabe11} that, in the case of weak c-f hybridization and thus a hypothetical magnetic QCP at pressure $p_{\rm QCP}$ higher than $p_{\rm{cr}}$ , it is possible that the VCO drives a first-order collapse of magnetism at $p_{\rm{c}}$ ($<p_{\rm{cr}}<p_{\rm QCP}$).
In this case, the peaks in $\rho_0$ and $A$ at $\sim 13$~GPa may be due to enhanced charge fluctuations due to competition between inter-site and Kondo–Yosida singlets, when $T_{\rm K}\sim T_1^{\rm max}$ becomes comparable to $T_{\rm N}$ \cite{hattori10}, which is indeed observed [see Fig. 2(a)].

(ii) When spin fluctuations are believed to mediate SC, optimal SC occurs close to the magnetic QCP and $T_{\rm{c}}$ does not exceed $\sim0.75$~K \cite{mathur98,yuan03}.
However in CeAg$_2$Si$_2$, optimal SC with $T_{\rm{c}}=1.25$~K occurs roughly 3~GPa above the magnetic $p_{\rm{c}}$, and partial SC persists up to $\sim21$~GPa, i.e, roughly 8~GPa above $p_{\rm{c}}$.

(iii) From a larger point of view, the magnetic phase diagram is different for each compound of the CeCu$_2$Si$_2$-HF family, while all other properties including SC are quite similar \cite{jaccard99,holmes04,seyfarth12,ren14,ren16,scheerer17}.

In summary, we have discovered HF SC in CeAg$_2$Si$_2$ and established its the $p$-$T$ phase diagram up to 22.5~GPa. The magnetic phase, whose transition line first increases with increasing pressure, suddenly collapses at $p_{\rm{c}}\sim13$~GPa. SC emerges close to $p_{\rm{c}}$, is optimal with $T_{\rm{c}}=1.25$~K at $p\sim16$~GPa, and persist over a huge pressure range of roughly 10~GPa.
Several features in the resistivity strongly suggest a pressure-induced VCO in proximity to optimal SC.
The crossover arises from a putative valence transition CEP at critical pressure $p_{\rm{cr}}\sim17$~GPa and temperature $T_{\rm{cr}}\sim-13$~K.
In particular, the isotherms of the normalized resistivity plotted versus the generalized distance from the CEP collapse on a single scaling function.
From these findings we emphasize that, besides spin fluctuations, valence fluctuations are a key ingredient in the low-temperature physics of CeAg$_2$Si$_2$.
New experiments with better resolution on the pressure scale will be essential to clarify the exact nature of the phase transition at $p_{\rm{c}}$ and of the SC pairing mechanism.

\section{Acknowledgments}

We acknowledge enlightening discussions with Shinji Watanabe and Kazumasa Miyake and technical support from Marco Lopez and Sebastian M\"{u}ller.

\end{document}